# Fast optical refocusing through multimode fiber bend using Cake-Cutting Hadamard encoding algorithm to improve robustness

CHUNCHENG ZHANG[1], ZHEYI YAO[1], ZHENGYUE QIN[1], GUOHUA GU[1], QIAN CHEN[1], ZHIHUA XIE[2], GUODONG LIU[2], XIUBAO SUI[1,*]

[1] *School of Electronic Engineering and Optoelectronic Technology, Nanjing University of Science and Technology, Nanjing 210094, China*

[2] *Key Lab of Optic-Electronic and Communication, Jiangxi Science and Technology Normal University, Nanchang 330013, China*

*sxb@njust.edu.cn

**Abstract:** Multimode fibres offer the advantages of high resolution and miniaturization over single-mode fibers in the field of optical imaging. However, multimode fibre's imaging is susceptible to perturbations of MMF that can lead to secondary spatial distortions in the transmitted image. Perturbations include random disturbances in the fiber as well as environmental noise. Here, we exploit the fast focusing capability of the Cake-Cutting Hadamard coding algorithm to counteract the effects of perturbations and improve the system's robustness. Simulation shows that it can approach the theoretical enhancement at 2000 measurements. Experimental results show that the algorithm can help the system to refocus in a short time when MMFs are perturbed. This research will further contribute to using multimode fibres in medicine, communication, and detection.

## 1. Introduction

When light is transmitted through a scattering medium, such as white paper, milk, and multimode fibre(MMF), the inhomogeneous refractive index of the transmission medium causes the incident light to be dislocated as bright and dark spots in the image plane[1]. To overcome this limitation, wavefront shaping techniques have been proposed [2]. This technique modulates the optical wavefront phase to help the incident light to form a well-defined focal point in the image plane [3]. Current methods for modulating wavefronts fall into three categories: optical phase conjugation [4, 5], transmission matrix measurement (TM) [6], and wavefront shaping based on the feedback of the target [7]. In recent years, wavefront shaping methods based on feedback have been explored in various fields, such as deep tissue microscopy [8, 9], optical capture [10], and endoscopy[11, 12].

Endoscopes have a wide range of applications in the medical field, which is of great interest to researchers. Traditional endoscopes consist of bundles of fibres integrated with multiple single-mode fibres (SMF), which significantly limits the further development of endoscopes in this field due to their large size and low resolution [13]. Today, technologies for preparing MMFs are mature. It can transmit multiple mutually independent channels in parallel [14]. With the development of MMF, small volume and high-resolution endoscopes have become possible [11]. However, the optical mapping between the incident and outgoing end of MMF is dislocated due to its mode coupling and dispersion [15]. As mentioned in the first paragraph, wavefront shaping techniques achieve superior optical focusing capabilities by numerical optimization, such as continuous sequence algorithms (CSA) [2], partitioning algorithms (PA) [16], genetic algorithms (GA) [17], simulated annealing algorithms (SAA) [18], etc. However, they relied heavily on feedback from the previous measurement to guide the subsequent optimization without considering the physical mechanisms behind the scattering process with perturbations. In practical scenarios, MMFs are dynamic scattering media. Once the scattering medium is perturbed, e.g. by high noise or changes in the medium itself beyond a particular state, the transport matrix of the medium will be completely altered, i.e. decorrelate, and then the algorithm will fail[19]. A new optimization process needs to be implemented. Moreover, the random perturbations that occur from time to time will disrupt the optical mapping between the incoming and outgoing ends, resulting in the phase of the wavefront from the previous iteration not matching and the focus point was weakening or even disappearing with the degree of perturbation [20].

Researchers have used experimental setups to overcome this instability to distinguish perturbations caused by MMFs [21]. On the other hand, adaptive optimization algorithms for wavefront shaping that can avoid dependence on previous iterations are also viable. Lai et al. developed a dynamic variational algorithm that escapes the firm reliance on previous optimizations and adapts to random perturbations of the MMFs [22]. Unfortunately, this algorithm only applies to binary wavefront shaping and is less robust in noisy environments than genetic algorithms. Yu et al. proposed an adaptive genetic algorithm that adapts variation according to the perturbation level of the MMFs to obtain better robustness [23]. However, the method has the following disadvantages: 1. Poor performance in refocusing after perturbation, requiring more time to exceed the enhancement of previous iterations or the theoretical optimum enhancement; 2. Tedious adjustment of parameters is required. In summary, developing a simple and efficient fast refocusing algorithm to attenuate or even ignore the effect of scattering medium's perturbations is a crucial goal of current research in this area. Hadamard matrices are used in various applications, including modern telecommunications, digital signal processing, and compressed sensing [24]. It has an important position in the field of image processing [25-29].

This paper uses a wavefront iteration algorithm by Cake-Cutting Hadamard matrix with an approach of four phase shift. Hadamard matrix is an orthogonal matrix, i.e. any two of its rows (columns) are orthogonal, and all elements have values of +1 or -1. It forms the combined weighing design, which implements optical multiplexing, dramatically accelerates the algorithm's convergence speed, and focus can acquire high enhancement. An imaging system can achieve a faster focusing capability than wavefront iteration algorithms such as GA, CSA, and SAA. Simulations and experiments demonstrate that the algorithm is not limited by the level of noise and strong disturbance. And it can achieve focus with high enhancements. The Cake-Cutting Hadamard algorithm (CHA) even gain theoretical enhancement with 2000 measurements. This approach will further advance the application of MMFs in image transmission and light focusing.

## 2. Principle

### 2.1 Wavefront shaping

The transmission of incident light through a scattering medium is deterministic. Mathematically, the scattering medium can be represented by the transmission matrix $t_{mn}$, which $m, n = 1, ..., N$. $t_{mn}$ fits the circular Gaussian distribution. The outgoing light field is:

$$E_m^{out} = \sum_n^N t_{mn} A_n e^{i\varphi_n} \tag{1}$$

$A_n$ and $\varphi_n$ are the amplitude and phase of the n-th controlled segment of the incident light. The equation represents that the outgoing light field $E_m^{out}$ of m-th segment's amplitude is formed by optical coherence of all segments in the incident light field, i.e. the sum of the transmission coefficients corresponding to the incident light multiplied by the $t_{mn}$. The focusing algorithm aims to find the best $\varphi_n$. Its corresponding light intensity is expressed as follows:

$$I_m^{out} = |E_m^{out}|^2 \tag{2}$$

$I_m^{out}$ is the light intensity of desired focus. The performance of focusing is defined as follows:

$$\eta = \frac{I_m^{out}}{\langle I_0 \rangle} \tag{3}$$

where $\langle I_0 \rangle$ is the average light intensity of the outgoing light and $\eta$ is the enhancement of the focusing point. The theoretical enhancement of the modulated phase method:

$$\eta_{ideal} = \pi(N-1)/4 \tag{4}$$

*2.2 Cake-Cutting Hadamard*

The most critical component of the Hadamard algorithm is the Hadamard matrix. Suppose the Hadamard matrix contains fewer connected regions (cake blocks), i.e. optimally reordered Hadamard bases. In this case, their corresponding measurements are likely to be higher. Therefore, the system gets more detection information, which can speed up the formation of a focus point. The specific methods are first taking each row of the larger Hadamard matrix and rearranging it into squares. Then, each square is arranged in descending order according to the number of connected areas it contains. Cake-cutting Hadamard matrix can be obtained by rearranging each row of the Hadamard matrix in this order, as shown in Fig. 1(a), and the cake-cutting Hadamard basis is shown in Fig.1(b). The optimal phase mask is then calculated according to the experimental procedure. The specific process is as follows:

The current optimum phase is calculated using the four phase method, which finds a balance between accuracy and speed of measurement compared to a three phase shift or five phase shift.

$$\varphi_{j,k} = \psi_{j-1}(x) + \frac{\left(\frac{k\pi}{2}\right)(H_j + 1)}{2}, \quad k = 0,1,2,3 \tag{5}$$

$$I_{m,k} = \left|\sum_m t_{mn} e^{i\varphi_{j,k}}\right|^2 \tag{6}$$

Where j is the same as $n$ in equation (1). $\psi_{j-1}$ denotes the optimal phase mask obtained at the n-th iteration, and $\frac{k\pi}{2}$ denotes the operation of phase-shift. $(H_j + 1)/2$ changed the values in the matrix to "0" and "+1" to meet the practical application. The four-step phase-shifting mechanism allows the determination of the retardation angle $\phi_n$ between the two phasors.

$$\phi_n = \text{Arg}[(I_{m,1} - I_{m,3}) + i(I_{m,2} - I_{m,4})] \tag{7}$$

Where Arg[·] The real part of the complex value is calculated. Finally, the optimal phase mask is updated:

$$\psi_n = \psi_{n-1} + \phi_n(H_n + 1)/2 \tag{8}$$

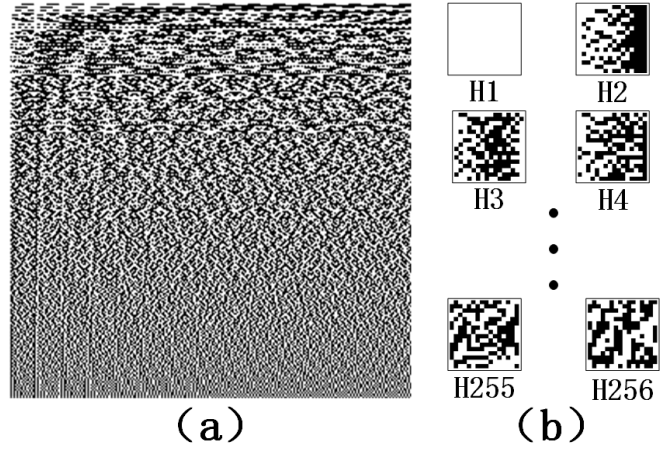

Fig. 1. Hadamard basis sort (a) 256th-order Cake-Cutting Hadamard; (b) A series of Hadamard bases is reshaped into two-dimensional matrices.

## 3. Simulation.

After the MMF modulates the incident light, the camera obtains a seemingly disordered scattering pattern, and the structure of this typical scattering optics system is shown in Fig.2(a). After optimizations of the wavefront algorithm, the liquid crystal phase modulator (SLM) constructs a relative optimum phase that helps the target light to form a focal point at a predetermined position in the image plane, as shown in Fig. 2(b). The modulation effect of the MMF on the incident light can be characterized as a matrix with a circularly symmetric complex Gaussian distribution. The Fresnel diffraction theory is then applied to simulate the spatial transmission and build the simulation environment. This section provides a preliminary verification of the performance of CHA.

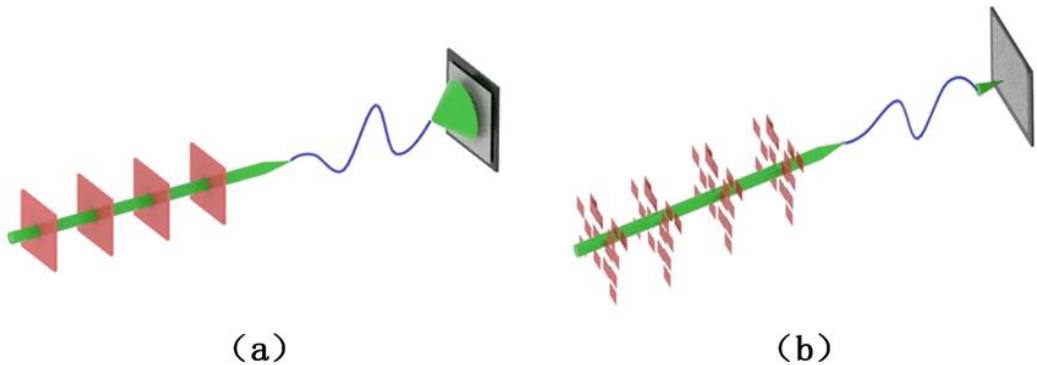

Fig.2. Schematic diagram of feedback-based wavefront shaping (a) light is scattered after passing through MMF (b) light is focused after passing through MMF after SLM modulation of the phase

We compare the performance of CHA with other well-known algorithms for different levels of perturbations and noises, including CSA, SAA, and GA. For all algorithms. The phase mask of size $N$ is set to 256, i.e., a phase mask of size $16 \times 16$. The phase takes a range of values from $0 - 2\pi$. The corresponding theoretical enhancement value is approximately 201. For CSA, the step size of the phase modulation is $\pi/10$. For the GA, the population size is 40; initial and final mutation rates are 0.1 and 0.001, respectively; and the decay factor of mutation is 200. The initial and final temperatures of SAA are set as 99 and 0.01, respectively. The length of its Markov chain is 20. The experimental results of each simulation were averaged over 100 independent optimizations.

*3.1 Impact of noise*

During the simulation, Gaussian white noise is added to the scattered light field to simulate the system and environmental noise. The without, low, medium and high noise are represented as $0 < I0 >$, $0.3 < I0 >$, $0.8 < I0 >$ and $1.5 < I0 >$, they can affect the performance of the focusing. Columns (a), (b), (c), and (d) of Fig. 3 clearly show the focusing ability of each algorithm in the case of different noise. Although the focusing performance of CSA is excellent without noise, it quickly reaches the neighborhood of the theoretical enhancement. However, as the noise level increases, the performance of CSA decreases sharply. Because the single-phase modulation method has a high initial convergence rate, but the corresponding feedback signal is easily drowned by noise. The GA performs well with robustness in all situations of noise. The enhancement of the focus always arrives near the ideal enhancement. Because the GA algorithm acts as a classical global optimization algorithm, owing to the consideration of all segments at the beginning and the effectiveness of the evolutionary strategy. The SAA algorithm mimics the principle of solid annealing by adding the metropolis principle to the greedy search, i.e. there is a certain probability of accepting a worse result, which helps the algorithm to jump out of the local optimum. However, the metropolis principle also makes the algorithm slower in searching for the global optimum. For the CHA, not only does it achieve the highest enhancement in all situations, but the focusing ability is hardly affected. It is worth noting that CHA already exceeds 50% of the ideal enhancement at 1000 measurements; at 2000 measurements, the enhancement value is close to 90%; and at around 4000 measurements, it has reached near the enhancement of theoretical value, which is the fastest than all other compared algorithms. Although CSA is close to CHA after 4000 measurements without noise, the enhancement is low at 2000 with only about 20% of the theoretical value. It converges at less than a quarter of the speed of CHA. The GA algorithm has an enhancement value of only 25% and 50% of the theoretical enhancement value at 2000 and 4000 measurements, respectively, which is much lower than the convergence speed of CHA.

*3.2 Effect of MMFs' disturbances*

In the imaging process of MMF, in addition to the noise from the instability of the optical system or the environment, the unstable state of the MMF can also significantly negatively contribute to the focusing effect. During the simulation, the transmission matrix (TM) representing the MMF is shifted at 15,000 measurements and filled with elements of the newly generated TM to achieve disturbances of the MMF [19]. Simulation are performed in the disturbances of 0%, 25%, 50%, 75% and 100%, respectively. Note, this paper defines the degree of disturbances in terms of the correlation between the scatter pattern with pre-and post-perturbation rather than the number of right-shifted matrix elements, which will be specified in the discussion section. Rows I, II, III, IV, and V of Fig. 3 clearly show the focusing capability of each algorithm at different degrees of disturbances. The enhancement corresponding to all algorithms decreases almost correspondingly after the occurrence of disturbances. The current optimal masks for these algorithms are produced based on the state of the MMF before the TM offset, which is closely related. For the post-offset TM, the correlation of the masks decreases, and the enhancement decreases accordingly. Each iteration of the CSA is relatively independent and therefore has a good refocusing capability for various levels of disturbances, recovering the enhancement of pre-perturbation except for noise. It has been shown in Section 3.1 that the GA has good immunity to noise. The enhancement can reach near the theoretical enhancement regardless of the effect of the noise. However, under varying degrees of random perturbation of the MMF, the refocusing capability is weak and never recovers enhancement of pre-disturbance because the GA generates new populations based on previous results.

The SAA uses the current phase mask as the initial population to continue optimization, which is also susceptible to disturbances. The CHA achieves the highest enhancement in all cases of disturbances and restores the enhancement of the focus to nearby theoretical value in a relatively short period. It is worth noting that regardless of the levels of disturbance, CHA has recovered the enhancement to around 80% of the theoretical enhancement at 1000 measurements. At 2000 measurements, the enhancement is close to 90%. And at 4000 measurements, it has reached near the theoretical value. Its convergence speed is the fastest of all the algorithms. Although the CSA has a slightly better-refocusing capability than the CHA after a low disturbance in the absence of noise, it has a much lower refocusing speed in the other disturbances. In practical applications, the presence of noise needs to be considered, and the final enhancement of the CSA is less effective. Regardless of the perturbations, the GA converges slowly, with the enhancement recovering only 50% of the theoretical enhancement (or at best less than 75% of the theoretical enhancement) after 5000 measurements because the GA algorithm requires a diverse population to find the global optimum solution. But, as the number of iterations increases, the variability of the GA becomes lower, and the population diversity decreases. It falls into a local optimum, leading to a decrease in convergence speed.

Firstly, the CHA has an excellent ability to control the focusing of incident light through MMF. It is insensitive to the intervention of noise and random disturbances. Finally, CHA always has good robustness and fast convergence capability.

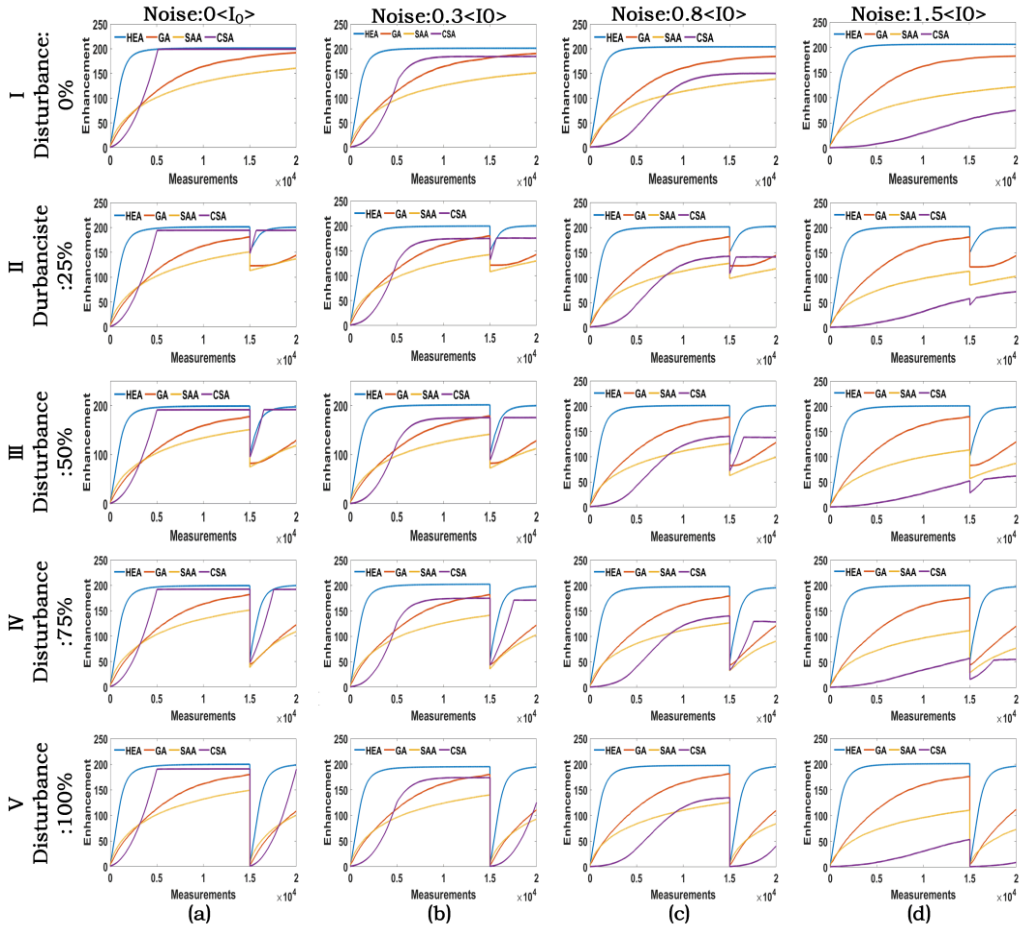

Fig.3. Simulation results of the effect of perturbations on the algorithm. Columns (a), (b), (c), and (d) represent $0 < I_0 >$, $0.3 < I_0 >$, $0.8 < I_0 >$ and $1.5 < I_0 >$ respectively. Rows Ⅰ, Ⅱ, Ⅲ, Ⅳ, and Ⅴ represent disturbances of 0%, 25%, 50%, 75% and 100%, respectively.

## 4. Experiment

*4.1 Experimental steps*

The experimental setup of the system is shown in Fig. 4. A continuous 532-nm laser (LCX-532S, Oxxius, 80 mW, polarization ratio: 100/1) is used as the light source. The light is expanded by the BE (GBE20-A-20 ×, Thorlabs, USA). Polarizers P is used to select the appropriate polarization state for the SLM (PLUTO-2-vis-096, HOLOEYE; pixel pitch: 8 μm). After being modulated and reflected by the SLM, the light is projected onto OBJ1 (RMS10X, 10× Thorlabs; NA:0.25) by a 4-f telescope system and then is coupled into an MMF. Mirror M is used to optimize the optical path. The scattered light from the MMF is collected by OBJ2 (RMS20X, 20×, Thorlabs; NA:0.4) and is finally detected by a CCD (BFS-U3-04S2M-CS, Point Gray; pixel pitch: 6.9 μm). There are 16×16 controlled segments involved in the illuminated area of the SLM which is a pure phase-type. A segment is a group of 20×20 adjacent SLM pixels. An MMF (DH-FMM200-FC-1C, NA: 0.5, length: 1m) is used to be the scattering sample in the paper. The central pixel of the CCD is chosen as the feedback position, and its surrounding 300×300 pixels are detected during the experimental optimization processes.

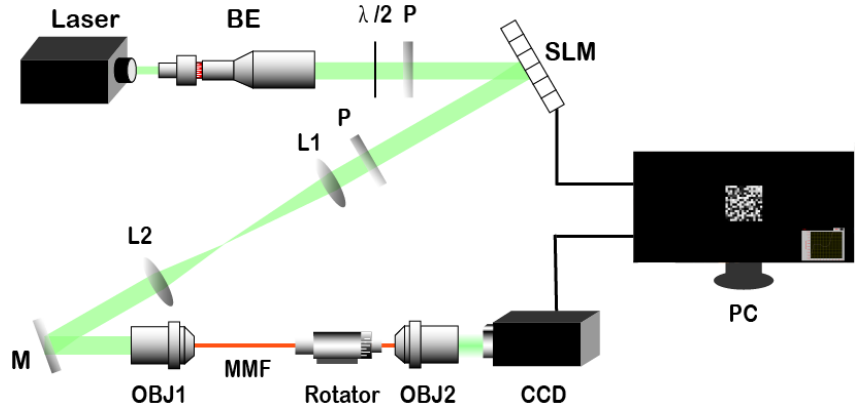

Fig.4. Chart of the optical system setup for experiments. BE: beam expander; L1, L2: lens; $\lambda/2$: half-wave plate; P: polarizer; M: mirror; OBJ1 and OBJ2: objectives; PC: a personal computer; SLM: spatial light modulator; MMF: multimode fiber.

*4.2 Focusing experiments*

In this section, all the parameters used in the experiments for the algorithms are consistent with the simulation. The results of single-point focusing using each algorithm are shown in Fig. 5. At the early stage of optimization (only 2000 measurements), the focusing results of the CHA, GA, CSA, and SAA are shown in Fig. 5(a), corresponding to enhancements of 54.8655, 41.05975, 11.5074, and 20.0954, respectively. The enhancement of CHA at 2000 measurements is already higher than the final enhancement of GA. CHA has the best convergence rate and the highest enhancement of all the algorithms. After 10000 measurements, all algorithms achieve successful focusing, as shown in Figs. 5(b-f), but the proposed CHA has the best performance of enhancement. The final enhancements of the CHA, GA, CSA, and SAA are 63.4927, 51.5941, 34.8024, and 21.1036, respectively. The final enhancement of CHA was 23.06%, 82.44%, and 200.86% higher than those of GA, CSA, and SAA, respectively. The experimental results show that the proposed CHA performs best for focusing in terms of not only convergence rate but also final enhancement.

The results of the final enhancement are lower than the simulation due to the noise of the optical system and the residual tuning amplitude of the LC-SLM. Other influencing factors include the coupling efficiency of the laser incident into the fiber and the calibration level of the optical system. CSA did not have a successful focus in the early stages probably because of the noise. Many bright dots appear in the background of the SAA, also because of the effect of noise. In contrast, both the CHA and the GA show good robustness.

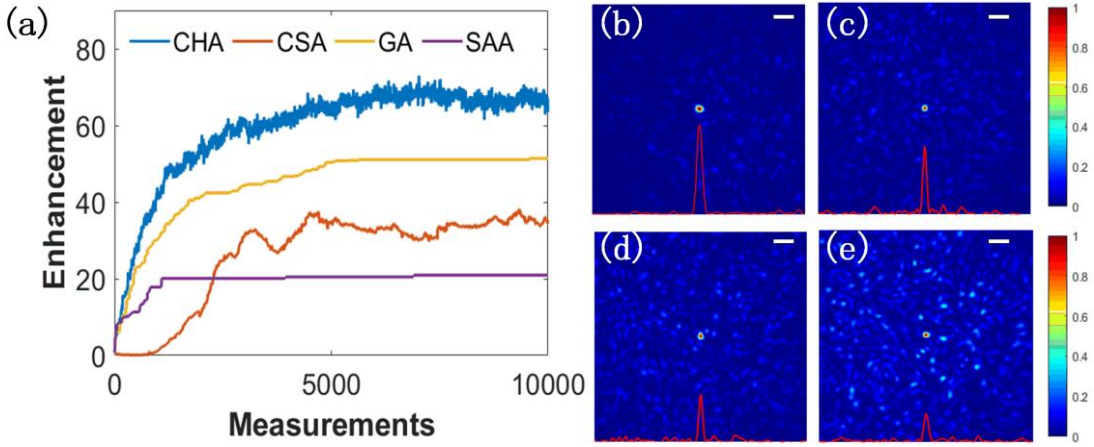

Fig.5. Experimental results after 20000 measurements. The scale bar is 138 μm, denoting a length of 20 CCD pixels. (a) Experimental enhancement curves of the CHA, GA, CSA, and SAA. (b-e) Results of CHA, GA, CSA, and SAA, corresponded to enhancements of 63.4927, 51.5941,34.8024, and 21.1036, respectively.

*4.3 Experiments of disturbance*

In the experiments, we use a fiber rotator to bend the MMF instead of the perturbation to which the MMF is subjected in real applications. This perturbation is quantitative, which ensures the fairness of the experiment. The rotation of the fiber rotator is 30 °, and then the scatter pattern is at or near disturbance: 100% [23]. The MMF was disturbed when the number of measurements was 8000, respectively.

The ability of CHA, GA, CSA, and SAA to refocus after MMF is disturbed, as shown in Fig. 6. At 8000 measurements, the focusing results of the CHA, GA, CSA, and SAA are shown in Figs. 6(a), corresponding to enhancements of 65.2996, 50.2795, 31.1794, and 24.6316, respectively. Then, the MMF is disturbed, the transmission matrix of the medium is changed, the correlation of the system is weakened, the enhancements of all algorithms immediately decrease, and the focus is destroyed. As shown in Fig. 6(a), their enhancements are close to 0, which indicates that they are almost in a completely de-correlated state before and after the disturbance. Then, the optimal phase masks currently obtained by all algorithms also lose their usefulness. They need to refocus on the state of the scattered pattern. After 2000 measurements i.e. 10000th measurement, all algorithms achieve successful focus, as shown in Figs. 6(b). The final enhancements of the CHA, GA, CSA, and SAA are 53.1347, 42.3522,17.0871, and 21.3970, respectively. The refocusing results of all algorithms are approximately the same as the simulation. The final enhancement of CHA was 25.46%, 210.96%, and 148.33% higher than those of GA, CSA, and SAA, respectively. It is worth noting that the enhancement of the proposed CHA after disturbance is higher than all other algorithms before disturbance in this paper. In conclusion, this experiment shows that CHA overcomes the GA population diversity, CSA depends on the long correlation time of the medium, and the low enhancement of SAA. It has a faster convergence speed and higher enhancement relative to GA, CSA, and SAA.

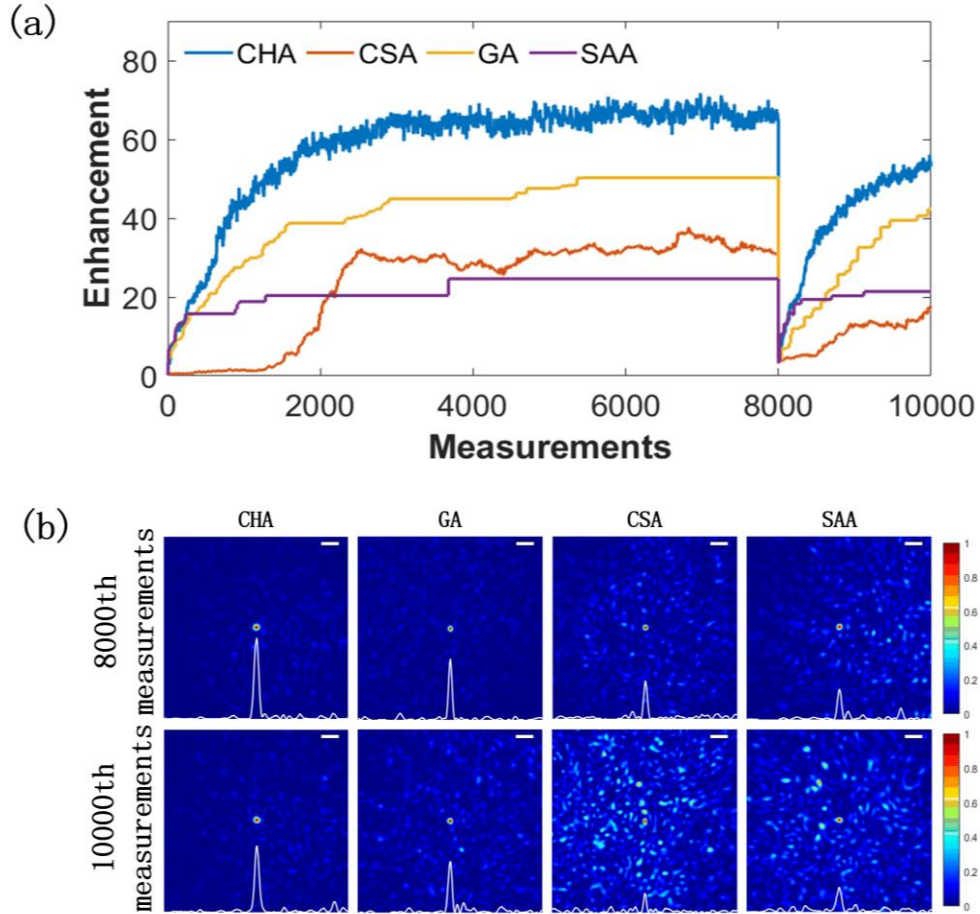

Fig.6. Experimental focusing enhancement curves of CHA, GA, CAS, and SAA when MMF is subjected to 30 °rotation. The rotation has occurred at 8000th measurements. Then, all algorithms continue to run 2000 measurements i.e. 10,000 measurements in total. (a) Experimental enhancement curves of the CHA, GA, CSA, and SAA. (b) Results of CHA, GA, CSA, and SAA, corresponded to enhancements of 53.2996, 50.3795,31.1794, and 24.6316 at 8000th measurements, respectively. Results of CHA, GA, CSA, and SAA, corresponded to enhancements of 53.1347, 42.3522,17.0871, and 21.3970 at 10000th measurements, respectively.

## 5. Discussion.

### 5.1 Analysis of noise

In the numerical simulations in Section 3.1, we found good robustness of the CHA to low, medium, and high noise disturbances. To further validate this performance of CHA, noise at $5 < I_0 >$ and $10 < I_0 >$ levels was added to simulations. The effect of noise on the stability of MMFs was tested, using the correlation of scattering patterns before and after the addition of noise as an evaluation metric. As shown in Fig. 7(a), the correlations for the $0.3 < I_0 >$, $0.8 < I_0 >$, $1.5 < I_0 >$, $5 < I_0 >$ and $10 < I_0 >$ noise were 98.3%, 90.0%, 74.0%, 31.4% and 16.4%, respectively. In autocorrelation imaging, the imaging system starts to decorrelate once the correlation of scatter falls below 50%. And the capability of imaging will be significantly reduced. If the correlation of scattering approaches 0, the imaging system is unable to image. Even if the scatter is highly decorrelated, the CHA algorithm still has good noise immunity, as shown in Fig. 7(b). No matter how much noise is applied to the imaging system, CHA can maintain the convergence rate and eventually reach a value near the theoretical enhancement.

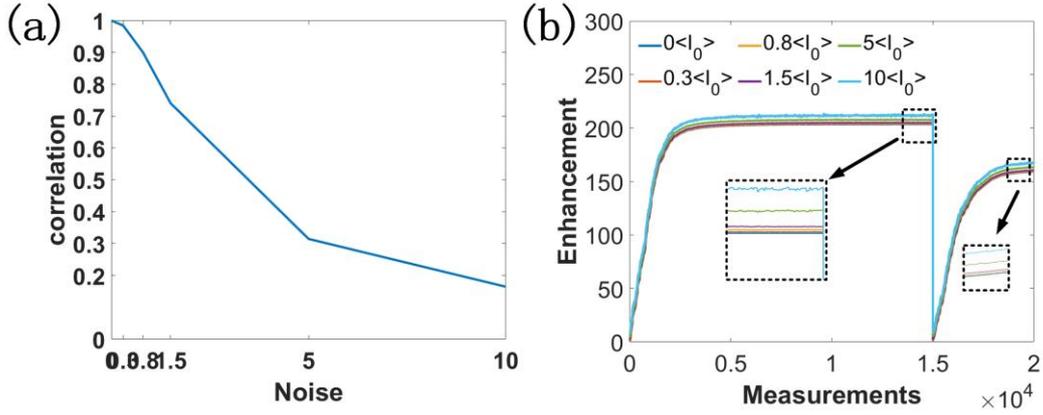

Fig.7. (a) Effects of levels of noise on the stability of MMFs (b) Performance of CHA for different levels of noise (with disturbance:100%)

### 5.2 Analysis of the disturbance

The section discusses the effect of the degree of MMF's disturbance on the focusing performance of the CHA. The correlation of speckle images before and after the disturbances is tested by gradually replacing new elements into the TM in steps of 1/256 of all elements in the TM. In the simulations, the correlation is 75% at an approximate disturbance of 34/256, 50% at an approximate disturbance of 75/256, and complete decorrelation at an approximate disturbance of 256/256, as shown in Fig. 8(a). After perturbation, the cases of 75%, 50%, 25%, and complete decorrelation correspond to enhancements of 150, 101, 53, and 3, respectively. The enhancement of the focus decreases by a value relative to the degree of decorrelation, which also demonstrates the accuracy of the scheme in simulated perturbation of MMFs. As shown in Fig. 8(b), after 1000 measurements, the enhancements are 182, 164, 146 and 117, recovering 16%, 31.5%, 46.5% and 57% of the theoretical enhancement, respectively. We can see that the stronger disturbance, the stronger the refocusing capability of the CHA due to the combined weighing design of Hadamard. After 3000 measurements, the enhancements reached more than 95% of the theoretical value. Because the combined "-1" and "+1" design helps the CHA to maintain a fast convergence rate at lower enhancements, random disturbances in the MMFs have less impact on the performance of CHA. It can be refocused within a short number of detections.

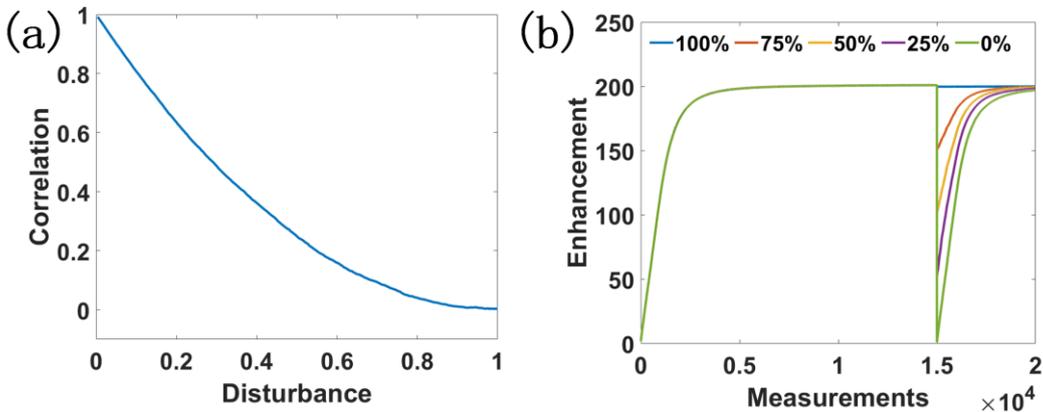

Fig.8. (a) Effect of varying perturbation levels on system relevance (b) Refocusing ability of CHA under different correlations of scatter pattern

Hadamard is unaffected by disturbances and noise. In the disturbance:100% and 1.5$<I_0>$ in Fig. 3, the refocusing capability of the CHA is shown in Table 1. With noise and random perturbation, the CHA

achieves 57.9%, 86.1%, 93.2%, 95.5%, and 96.5% of the theoretical enhancement at 1000, 2000, 3000, 4000, and 5000 detections, respectively, with a slight and almost negligible decrease in refocusing capability.

Table 1 Refocusing performance of CHA at disturbances:100% and 1.5<I0>.

| Number | 1000 | 2000 | 3000 | 4000 | 5000 |
| --- | --- | --- | --- | --- | --- |
| Enhancement | 117 | 174 | 188 | 193 | 195 |
| Refocusing | 57.9% | 86.1% | 93.0% | 95.5% | 96.5% |

## 5. Conclusion

In summary, we have demonstrated that CHA can converge quickly and effectively cope with random perturbations in MMFs, achieving good focusing results. Firstly, the algorithm's performance is unaffected by any level of noise. Then, CHA can refocus quickly in the face of disturbances due to its excellent convergence capability. Even after disturbance of an MMF, where the scatter pattern is entirely decorrelated and missing a focus point, the CHA can still recover more than 85% of the theoretical enhancement after 2000 measurements. To the best of our knowledge, the CHA has the fastest speed of refocusing on an iterative optimization method with wavefront feedback after perturbations of MMF. Despite these advantages, the performance of CHA is open to debate if MMF is subjected to sustained and frequent perturbations. To address this issue, we will consider replacing the SLM with a digital microlens device (DMD) in subsequent work because the DMDs we currently have access to have refresh rates of up to 10 WHz, which is expected to wholly or partially solve this problem. Our experimental results push the imaging boundaries of MMF.

**Declaration of interests**

The authors declare that they have no known competing financial interests or personal relationships that could have appeared to influence the work reported in this paper.

**Acknowledgments**

This work was supported by the National Natural Science Foundation of China ((No. 62105152), Key Research & Development programs in Jiangsu China(Grant no. BE2018126), Fundamental Research Funds for the Central Universities (Grant NO. JSGP202202, 30919011401，30920010001)，Leading Technology of Jiangsu Basic Research Plan (BK20192003), The Postgraduate Research & Practice Innovation Program of Jiangsu Province(KYCX22_0411), The Open Foundation of Key Lab of Optic-Electronic and Communication of Jiangxi Province (NO.20212OEC002).